\documentclass[sigconf,10pt]{acmart}
\usepackage{enumitem}
\usepackage{graphicx}
\graphicspath{ {./images/} }
\AtBeginDocument{%
  \providecommand\BibTeX{{%
    \normalfont B\kern-0.5em{\scshape i\kern-0.25em b}\kern-0.8em\TeX}}}

\acmYear{2022}\copyrightyear{2022}
\setcopyright{rightsretained}
\acmConference[ACM MobiCom '22]{The 28th Annual International Conference on Mobile Computing and Networking}{October 17--21, 2022}{Sydney, NSW, Australia}
\acmBooktitle{The 28th Annual International Conference on Mobile Computing and Networking (ACM MobiCom '22), October 17--21, 2022, Sydney, NSW, Australia}
\acmPrice{15.00}
\acmDOI{10.1145/3495243.3558752}
\acmISBN{978-1-4503-9181-8/22/10}

\begin{document}

\title{DIY-IPS: Towards an Off-the-Shelf Accurate Indoor Positioning System}



\author{Riccardo Menon}
\affiliation{%
  \institution{The University of Sydney}
  \streetaddress{1 Th{\o}rv{\"a}ld Circle}
  \city{Sydney }
  \country{Australia}}
\email{rmen8914@uni.sydney.edu.au}
\orcid{0000-0003-4538-9710}

\author{Abdallah Lakhdari}
\affiliation{%
  \institution{The University of Sydney}
  \city{Sydney }
  \country{Australia}}
\email{abdallah.lakhdari@sydney.edu.au}
\orcid{0000-0001-8005-1534}

\author{Amani Abusafia}
\affiliation{%
  \institution{The University of Sydney}
  \city{Sydney }
  \country{Australia}}
\email{amani.abusafia@sydney.edu.au}
\orcid{0000-0001-9159-6214}

\author{Qijun He}
\affiliation{%
  \institution{The University of Sydney}
  \city{Sydney }
  \country{Australia}}
\email{qihe9584@uni.sydney.edu.au}
\orcid{0000-0003-0933-9054}

\author{Athman Bouguettaya}
\affiliation{%
  \institution{The University of Sydney}
  \city{Sydney }
  \country{Australia}}
\email{athman.bouguettaya@sydney.edu.au}
\orcid{0000-0003-1254-8092}

\renewcommand{\shortauthors}{Riccardo and Abdallah, et al.}

\begin{abstract}

We present \textit{DIY-IPS - Do It Yourself - Indoor Positioning System}, an open-source real-time indoor positioning mobile application. \textit{DIY-IPS} detects users' indoor position by employing\textit{ dual-band RSSI} fingerprinting of available WiFi access points. The app can be used, without additional infrastructural costs, to detect users' indoor positions in real time. We published our app as an \textit{open source} to save other researchers time recreating it. The app enables researchers/users to  (1) collect indoor positioning datasets with a ground truth label, (2) customize the app for higher accuracy or other research purposes (3) test the accuracy of modified methods by live testing with ground truth. We ran preliminary experiments to demonstrate the effectiveness of the app.
\vspace{-10pt}


\end{abstract}

\ccsdesc[500]{Information systems~Mobile information processing systems}
\ccsdesc[500]{Information systems~Global positioning systems\vspace{-10pt}}

\keywords{indoor localization, indoor positioning, indoor datasets, indoor navigation, smartphones}


\maketitle
\vspace{-10pt}
\section{Introduction}
\vspace{-2pt}
Indoor positioning refers to detecting an object's location in an enclosed environment, e.g., buildings or malls \cite{zafari2019survey}. Indoor location is considered the most important context information because of its necessity in various real-world applications, such as path navigation \cite{chen2018easygo}, location-based services \cite{lakhdari2021proactive,yao2022wireless,lakhdari2021fairness}, and marketing \cite{farahsari2022survey,Amani2022QoE}. Existing literature proposed several approaches and technologies to detect indoor locations \cite{chintalapudi2010indoor}. However, they are either expensive or have low accuracy \cite{zafari2019survey,farahsari2022survey,abbas2019wideep}.
\looseness=-1 


One of the prominent indoor positioning solutions is the use of the Received Signal Strength (RSS) of WiFi Access Points (APs) \cite{cao2021fingerprint,yang2015wifi}. The use of WiFi offers a ubiquitous and cost-effective solution due to the high availability of WiFi services in indoor environments nowadays \cite{zafari2019survey}. Additionally, RSS values can be easily captured by devices with built-in WiFi chips, e.g., smartphones \cite{chintalapudi2010indoor}. In general, this approach relies on estimating the distance between the device and the available access points by assessing the strength of their signals \cite{cao2021fingerprint}. The accuracy of using WiFi is low due to the impact of the ambient conditions such as the crowds and other devices \cite{roy2022survey}. Thus, a study proposed the use of dual-band WiFi for indoor positioning based on Gaussian Process Regression (GPR) and the k-nearest neighbor (KNN) \cite{cao2021fingerprint}. However, their accuracy is low compared to other existing expensive solutions \cite{caso2019vifi}.\looseness=-1

\begin{figure}
    \centering
    \setlength{\abovecaptionskip}{0pt}
    \setlength{\belowcaptionskip}{-15pt}
    \includegraphics[width=\linewidth]{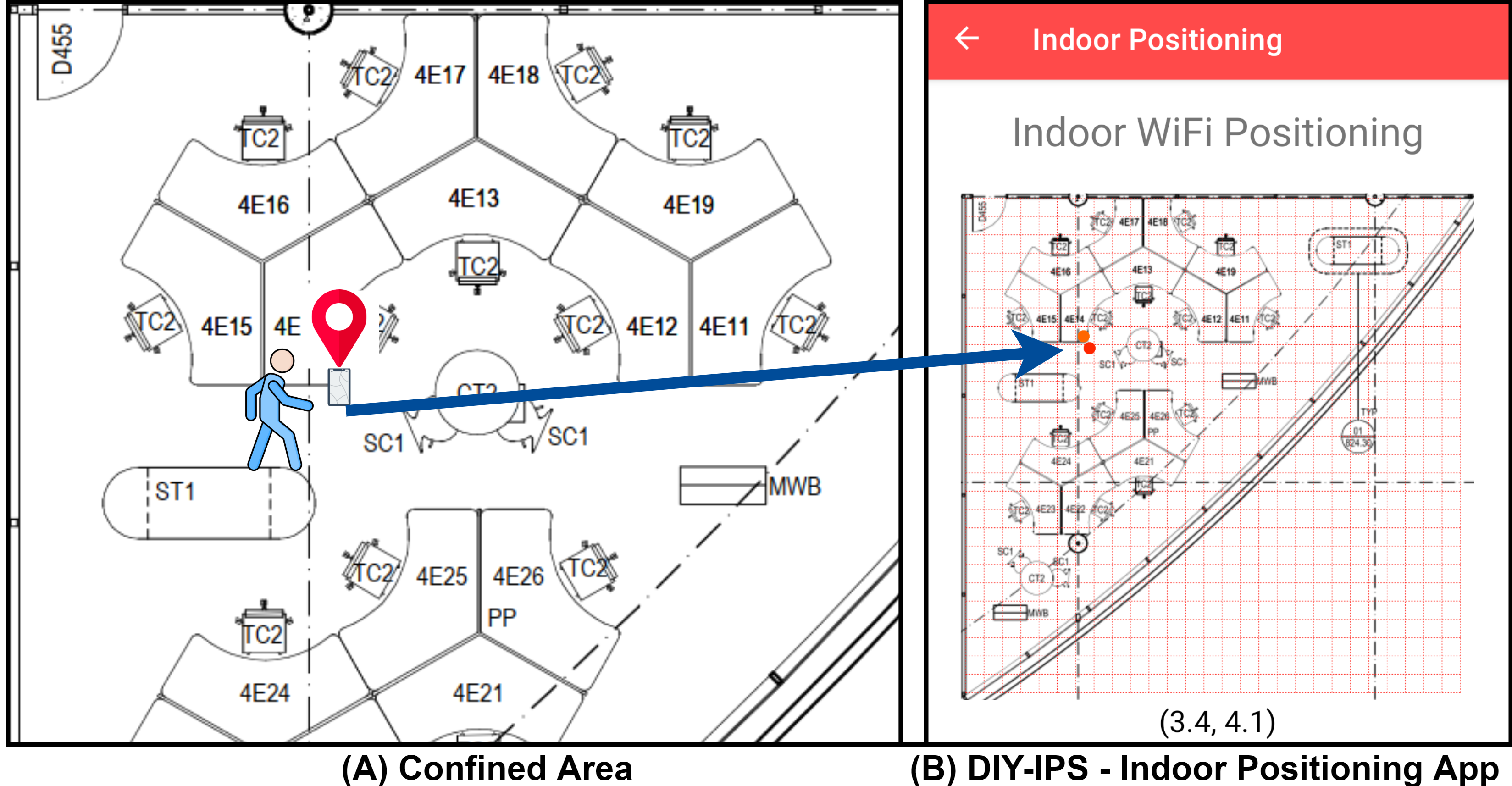}
    \caption{Example of Using \textit{DIY-IPS} Application}
    \label{fig:context}
\end{figure}

We propose \textit{DIY-IPS}, an open-source real-time indoor positioning mobile application that uses \textit{wifi dual band} to estimate indoor location. \textit{DIY-IPS} doesn't require any infrastructure installation to be used. Thus, it doesn't require any additional costs. In detail, our application uses the existing WiFi access points in a confined area to : (1) collect RSS data, (2) train a model (3) estimate the location using the model. Hence, inexperienced users (e.g., a mall owner)  may  use the  app  for  different applications including indoor navigation (Fig. \ref{fig:context}). Additionally, we provide \textit{DIY-IPS} as  open-source application to save other researchers time in recreating the app and enable them to: (1) customize the model to try new models that may offer higher accuracy. (2) test the new model's accuracy by comparing it to the ground truth. The app's source code is available at https://github.com/SensorsCloudsServicesLab/DIY\_IPS-Indoor-Positioning-System.\looseness=-1

\vspace{-3pt}
\section{Overview}
Our application consists of two phases: (1) An offline phase where the training data is collected, and the model is trained. (2) An online phase where the location is estimated. Both phases are represented by five buttons in our app (Fig.\ref{fig:app}).\looseness=-1 
\vspace{-10pt}
\subsection{Offline Phase}
The offline phase is represented by three steps/buttons:
\begin{enumerate}[ noitemsep,nosep,leftmargin=10pt,labelsep=5pt,itemindent=10pt, labelwidth=*]
 \item \textbf{Collecting the RSSI Data:} The collection of the RSSI fingerprint dataset must be done prior to using the live indoor localization or testing the accuracy. The procedure for collecting the dataset for a confined area, e.g., a room, is: (A) divide the room  into a set of reference points (B) At each reference point,  collect  RSSI values from four directions (N, E, S, W). 
After opening the "RSSI collect data" section of the app, the user's true position can be entered as a reference point. After clicking a button, the app automatically collects RSSI samples from available access points, and uploads the data to a server. Each RSSI fingerprint sample consists of (A) the  coordinates of the reference point. (B) the angle that the phone was facing relative to North. (C) RSS values of all dual-band WiFi Access points.
    \item \textbf{Processing the data distribution:} In this step, the collected data will be retrieved from the server. Then, a normal distribution is calculated at each reference point for each direction and WiFi access point combination.
    
    \item \textbf{Gaussian Process Regression (GPR):} Gaussian Process Regression is applied to generate a complete fingerprint of normal distributions for each direction.
    

\end{enumerate}

\vspace{-10pt}
\subsection{Online Phase}
\begin{figure}[!t]
\centering
    \begin{minipage}{0.48\linewidth}
    \centering
    \setlength{\abovecaptionskip}{1pt}
    \setlength{\belowcaptionskip}{-15pt}
    \includegraphics[width=\linewidth]{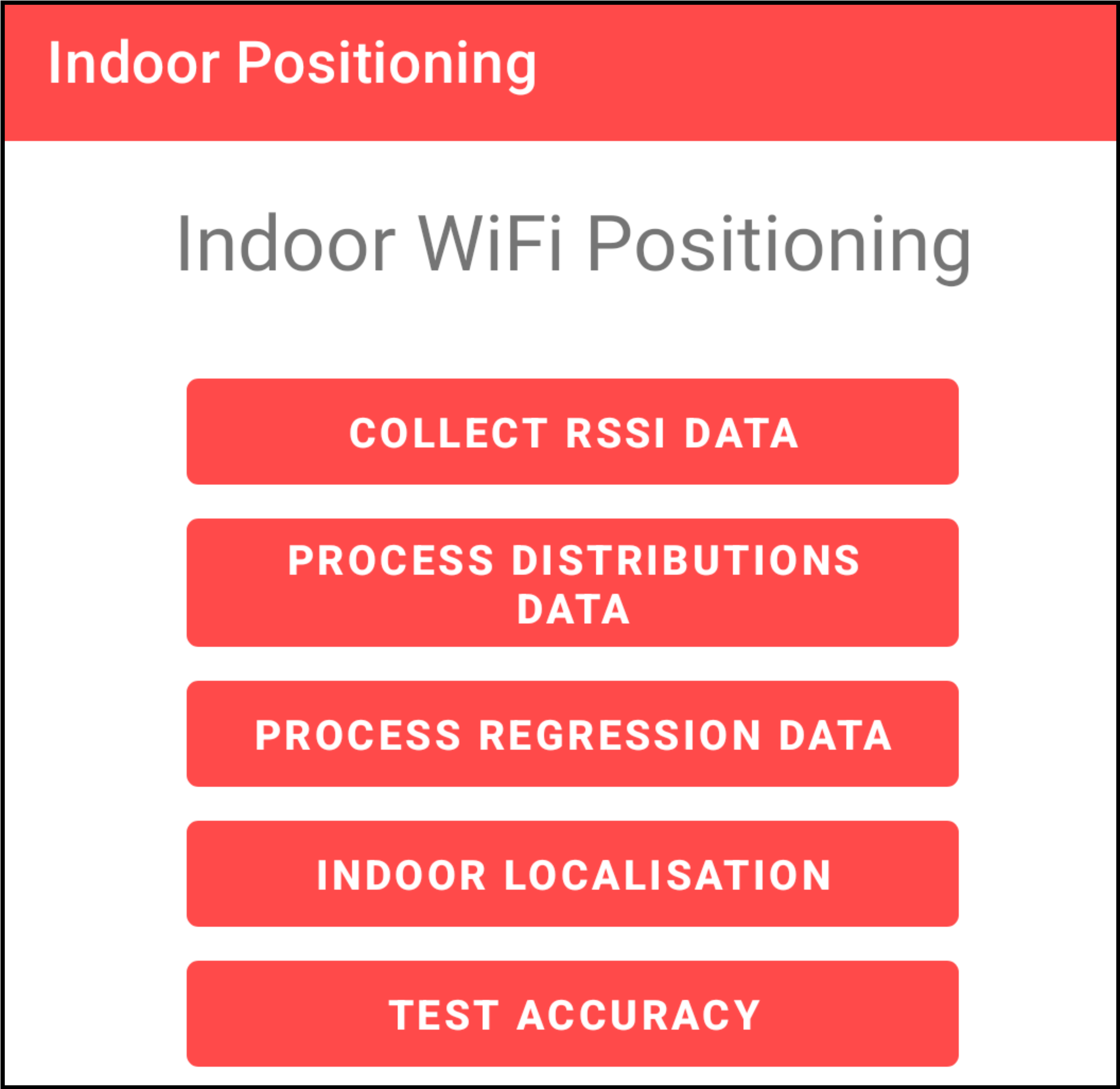}
    \caption{\textit{DIY-IPS} App}
    \label{fig:app}
    \end{minipage}
    \hfill
    \begin{minipage}{.48\linewidth}
    \centering
    \setlength{\abovecaptionskip}{1pt}
    \setlength{\belowcaptionskip}{-15pt}
    \includegraphics[width=0.95\linewidth]{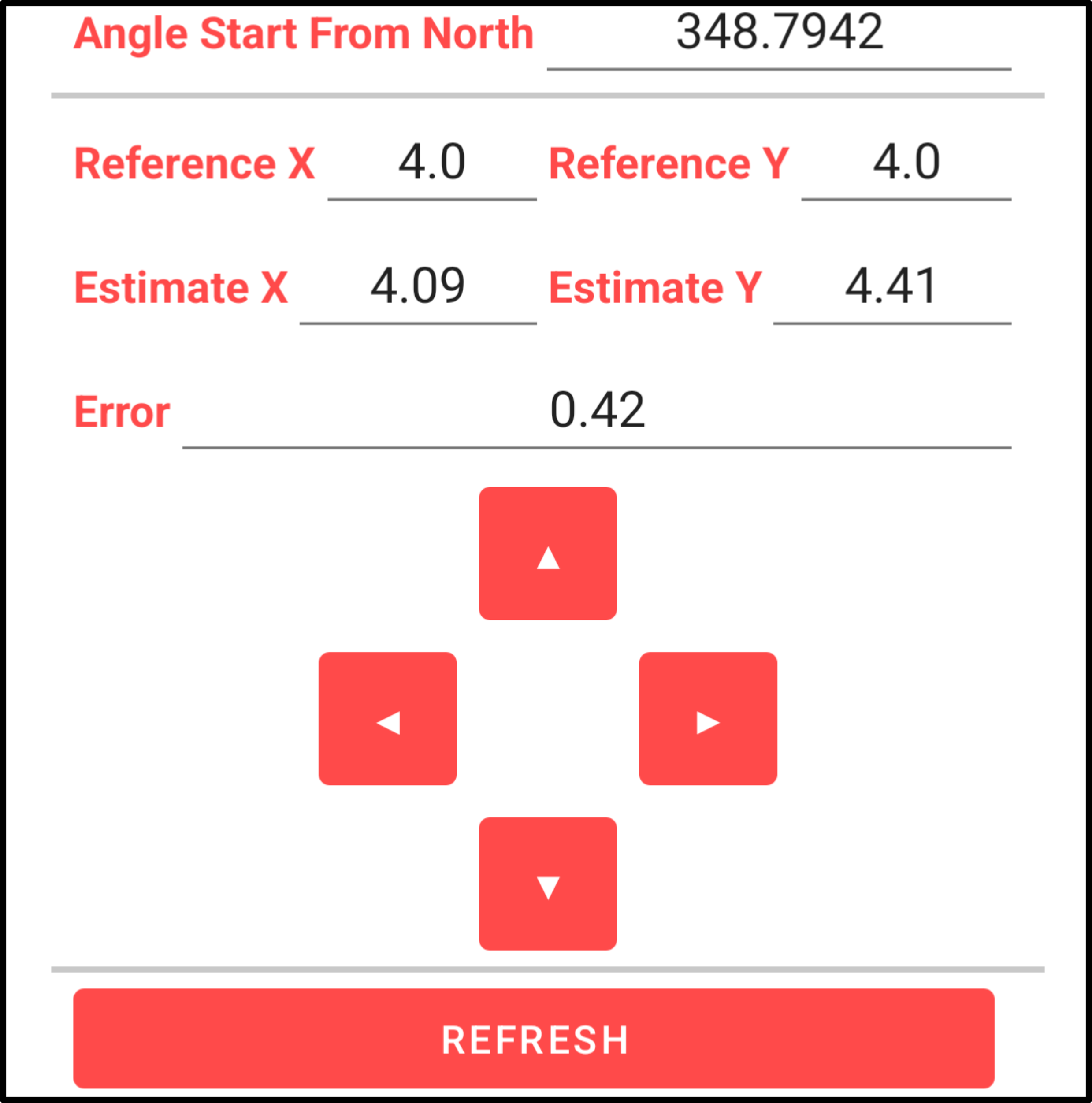}
    \caption{Testing Accuracy}
    \label{fig:test}
    \end{minipage}
\end{figure}

The online phase is represented by two steps/buttons:
\begin{itemize}[ noitemsep,nosep,leftmargin=10pt,labelsep=1pt,itemindent=0pt, labelwidth=*]
  \item \textbf{Indoor localization:} This button allows the user to visualise their location in real-time (Fig.\ref{fig:context}).
    \item \textbf{Accuracy Testing:} The Accuracy Testing button allows the user to input the ground truth. Then, the app will estimate the location using our model (Fig.\ref{fig:test}). Then, the error is calculated using the Euclidean distance between the ground truth and the estimated location. Finally, the mean error will be recorded locally.
\end{itemize}
\vspace{-10pt}

 \vspace{-5pt}
\subsection{Evaluation}
\begin{figure}[!t]
    \centering
        \setlength{\abovecaptionskip}{1pt}
    \setlength{\belowcaptionskip}{-15pt}
    \includegraphics[width=0.6\linewidth]{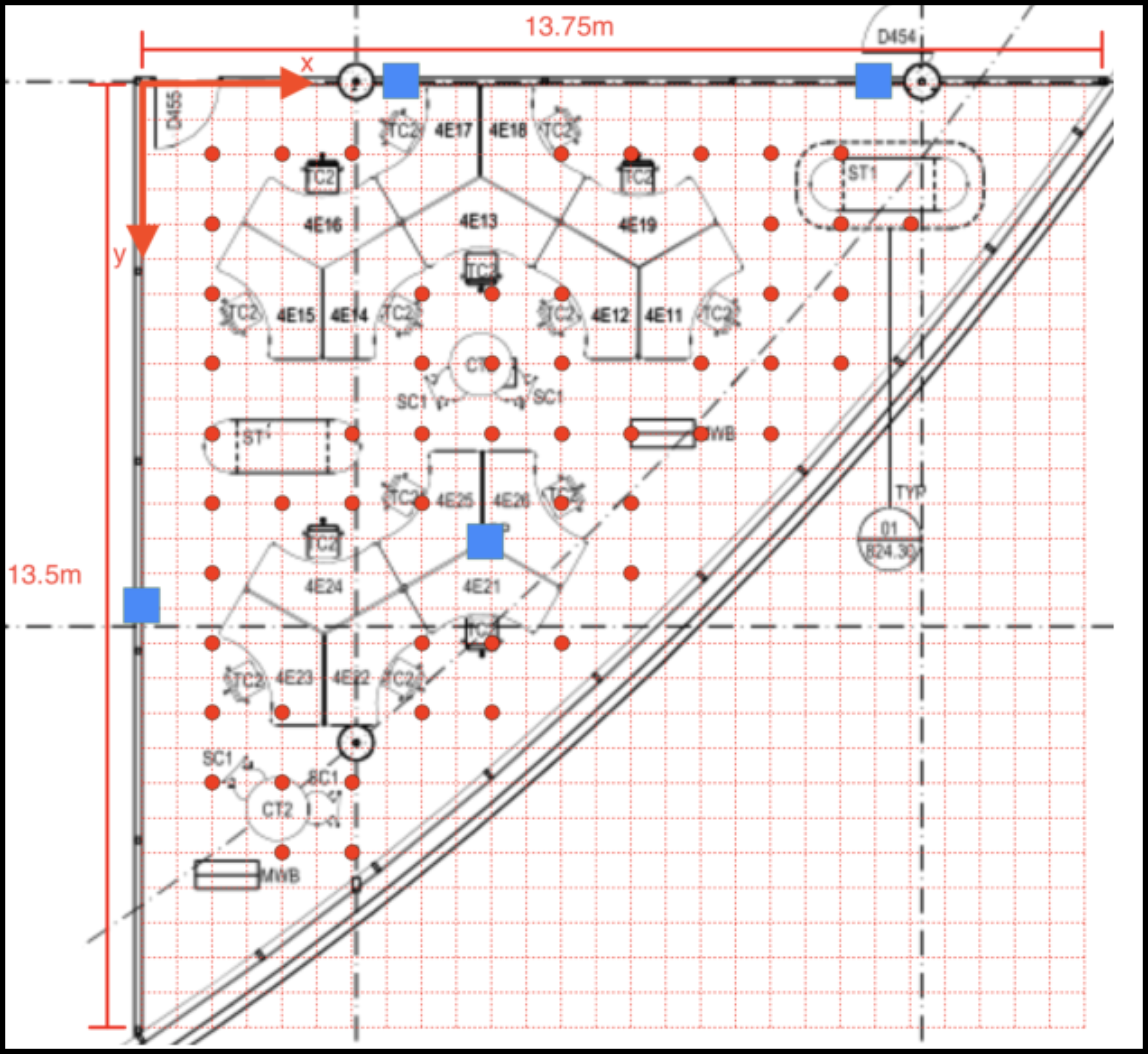}
    \caption{Map of The Room}
    \label{fig:room}
\end{figure}

We ran an experiment to test the accuracy of our model. we have used a room with  13.75m x 13.5m dimensions and created 54 reference points (Fig.\ref{fig:room}). The distance between each of them is 1m. The blue boxes in the figure represent the WiFi access points. We collected 43200 RSSI Samples. The mean error is 3.32 m, and the standard deviation is 1.93 m. The accuracy may be improved using a different model.\looseness=-1 




\vspace{-10pt}
\section{Demo Setup}
Our demo will exhibit the \textit{DIY-IPS} app positioning capabilities. Additionally, we will display a recorded video of the entire process of using the app to collect, train, test, and detect a user's location in real-time. The video is published at this link: youtu.be/Mib0leMLBIo. To demonstrate the efficiency of our model, we will run our framework using our collected RSSI database on a laptop. We can train our app to present it if we are given a map and early access to the conference hall. Training the app will enable the audience to use our smartphones during the demo session to show their locations in real time. We will use our laptops and smartphones, but we need a desk, power outlets, and internet ports. We will set up our Wi-Fi access point for the smartphones used for demonstration. The whole setup process should take less than two hours.\looseness=-1
\vspace{-8pt}
\section*{Acknowledgment}
This research was partly made possible by  LE220100078 and LE180100158 grants from the Australian Research Council. The statements made herein are solely the responsibility of the authors.




\bibliographystyle{unsrt}
\bibliography{main}

\end{document}